\documentstyle[epsfig,12pt]{article}
\begin{document}

\catcode`@=11
\long\def\@caption#1[#2]#3{\par\addcontentsline{\csname
  ext@#1\endcsname}{#1}{\protect\numberline{\csname
  the#1\endcsname}{\ignorespaces #2}}\begingroup
    \small
    \@parboxrestore
    \@makecaption{\csname fnum@#1\endcsname}{\ignorespaces #3}\par
  \endgroup}
\catcode`@=12
\newcommand{\newc}{\newcommand}
\newc{\gsim}{\lower.7ex\hbox{$\;\stackrel{\textstyle>}{\sim}\;$}}
\newc{\lsim}{\lower.7ex\hbox{$\;\stackrel{\textstyle<}{\sim}\;$}}
\newc{\gev}{\,{\rm GeV}}
\newc{\mev}{\,{\rm MeV}}
\newc{\ev}{\,{\rm eV}}
\newc{\kev}{\,{\rm keV}}
\newc{\tev}{\,{\rm TeV}}
\newc{\mz}{m_Z}
\newc{\mpl}{M_{\rm{Planck}}}
\newc{\chifc}{\chi_{{}_{\!F\!C}}}
\newc\order{{\cal O}}
\newc\CO{\order}
\newc\CL{{\cal L}}
\newc\CY{{\cal Y}}
\newc\CH{{\cal H}}
\newc\CM{{\cal M}}
\newc\CF{{\cal F}}
\newc\CD{{\cal D}}
\newc\CN{{\cal N}}
\newc{\eps}{\epsilon}
\newc{\re}{\mbox{Re}\,}
\newc{\im}{\mbox{Im}\,}
\newc{\invpb}{\,\mbox{pb}^{-1}}
\newc{\invfb}{\,\mbox{fb}^{-1}}
\newc{\yddiag}{{\bf D}}
\newc{\yddiagd}{{\bf D^\dagger}}
\newc{\yudiag}{{\bf U}}
\newc{\yudiagd}{{\bf U^\dagger}}
\newc{\yd}{{\bf Y_D}}
\newc{\ydd}{{\bf Y_D^\dagger}}
\newc{\yu}{{\bf Y_U}}
\newc{\yud}{{\bf Y_U^\dagger}}
\newc{\ckm}{{\bf V}}
\newc{\ckmd}{{\bf V^\dagger}}
\newc{\ckmz}{{\bf V^0}}
\newc{\ckmzd}{{\bf V^{0\dagger}}}
\newc{\X}{{\bf X}}
\newc{\bbbar}{B^0-\bar B^0}
\def\bra#1{\left\langle #1 \right|}
\def\ket#1{\left| #1 \right\rangle}
\newc{\sgn}{\mbox{sgn}\,}
\newc{\m}{{\bf m}}
\newc{\msusy}{M_{\rm SUSY}}
\newc{\munif}{M_{\rm unif}}
\newc{\slepton}{{\tilde\ell}}
\newc{\Slepton}{{\tilde L}}
\newc{\sneutrino}{{\tilde\nu}}
\newc{\selectron}{{\tilde e}}
\newc{\stau}{{\tilde\tau}}
%
%
\def\NPB#1#2#3{Nucl. Phys. {\bf B#1} (19#2) #3}
\def\PLB#1#2#3{Phys. Lett. {\bf B#1} (19#2) #3}
\def\PLBold#1#2#3{Phys. Lett. {\bf#1B} (19#2) #3}
\def\PRD#1#2#3{Phys. Rev. {\bf D#1} (19#2) #3}
\def\PRL#1#2#3{Phys. Rev. Lett. {\bf#1} (19#2) #3}
\def\PRT#1#2#3{Phys. Rep. {\bf#1} (19#2) #3}
\def\ARAA#1#2#3{Ann. Rev. Astron. Astrophys. {\bf#1} (19#2) #3}
\def\ARNP#1#2#3{Ann. Rev. Nucl. Part. Sci. {\bf#1} (19#2) #3}
\def\MPL#1#2#3{Mod. Phys. Lett. {\bf #1} (19#2) #3}
\def\ZPC#1#2#3{Zeit. f\"ur Physik {\bf C#1} (19#2) #3}
\def\APJ#1#2#3{Ap. J. {\bf #1} (19#2) #3}
\def\AP#1#2#3{{Ann. Phys. } {\bf #1} (19#2) #3}
\def\RMP#1#2#3{{Rev. Mod. Phys. } {\bf #1} (19#2) #3}
\def\CMP#1#2#3{{Comm. Math. Phys. } {\bf #1} (19#2) #3}
\relax
%
%
%
\def\beq{\begin{equation}}
\def\eeq{\end{equation}}
\def\bea{\begin{eqnarray}}
\def\eea{\end{eqnarray}}
%
%
%
\newc{\ie}{{\it i.e.}}          \newc{\etal}{{\it et al.}}
\newc{\eg}{{\it e.g.}}          \newc{\etc}{{\it etc.}}
\newc{\cf}{{\it c.f.}}
\def\smuon{{\tilde\mu}}
\def\neut{{\tilde N}}
\def\char{{\tilde C}}
\def\bino{{\tilde B}}
\def\wino{{\tilde W}}
\def\higgsino{{\tilde H}}
\def\sneut{{\tilde\nu}}
%
%
%
%
\def\slash#1{\rlap{$#1$}/} 
\def\Dsl{\,\raise.15ex\hbox{/}\mkern-13.5mu D} 
\def\delsl{\raise.15ex\hbox{/}\kern-.57em\partial}
\def\Ksl{\hbox{/\kern-.6000em\rm K}}
\def\Asl{\hbox{/\kern-.6500em \rm A}}
\def\Qsl{\hbox{/\kern-.6000em\rm Q}}
\def\gradsl{\hbox{/\kern-.6500em$\nabla$}}
%
%
%
\def\bar#1{\overline{#1}}
\def\vev#1{\left\langle #1 \right\rangle}
%

\begin{titlepage} 
\begin{flushright}
{DCP-08-02}\\
SHEP-08-24\\
\today
\end{flushright}
\vskip 2cm
\begin{center}
{\Large\bf{Lepton Flavour Violating Heavy Higgs Decays 
\\[0.25cm]
Within the $\nu$MSSM and Their Detection at the LHC }}
\vskip 1.25cm
{{\large
 J.L. Diaz-Cruz$^{(a,b)}$, Dilip Kumar Ghosh$^{(c)}$ and S. Moretti$^{(d)}$} \\
~~ \\
$^{(a)}${\it C.A. de Particulas, Campos y Relatividad, FCFM-BUAP\\
Puebla, Pue.,  72570, M\'exico}\\ 
~~ \\
$^{(b)}${\it DUAL CP Institute of High Energy Physics\\ 
Puebla, Pue.,  72570, M\'exico}\\ 
~~ \\
$^{(c)}${\it Theoretical Physics Division\\
Physical Research Laboratory\\
Navrangpura, Ahmedabad--380 009, India}\\     
~~ \\
$^{(d)}${\it School of Physics and Astronomy, University of Southampton \\
Highfield, Southampton SO17 1BJ, UK}\\[0.1truecm]
}

\end{center}
\vskip 1.cm

\begin{abstract}

\noindent
Within the $\nu$MSSM, a Minimal Supersymmetric neutrino See-saw Model, 
Lepton Flavour Violating  Higgs couplings are strongly enhanced 
at large $\tan\beta$ ($\gsim30$), which can lead to 
BR$(H^0/A^0 \to \tau\mu) \simeq O(10^{-4})$, for $M_{H^0/A^0}\gsim 160$ GeV.
Enhancements on the production of Higgs bosons, through the gluon 
fusion mechanism, $gg\to H^0/A^0$, and the associated production channel
$gg,q\bar q\to b\bar bH^0/A^0$, whose rates  grow with $\tan\beta$,
as well as the mass degeneracy that occurs between the $H^0$ and $A^0$
states in this regime, also contribute to further the possibilities to
detect a heavy Higgs signal into $\tau\mu$ pairs. We show that 
the separation of $\tau\mu$ Higgs events from the background at the
upcoming CERN Large Hadron Collider could be done for Higgs masses up to 
about 600 GeV for 300 fb$^{-1}$ of luminosity,
for large $\tan\beta$ values. However, even with as 
little as 10  fb$^{-1}$ one can probe $H^0/A^0$ masses up to 400 GeV or
so,
if  $\tan\beta=60$. Altogether, these processes then
provide a new Higgs discovery mode as well 
as an independent test of flavour physics.
\end{abstract}

\end{titlepage}

\setcounter{footnote}{0}
\setcounter{page}{1}
\setcounter{section}{0}
\setcounter{subsection}{0}
\setcounter{subsubsection}{0}

\section{Introduction}
\label{sect:intro}

It is possible that some of the most exciting times in particle physics
will come soon, as the upcoming Large Hadron Collider
(LHC) will enable us to probe
the mechanism of Electro-Weak Symmetry Breaking (EWSB). 
The standard picture contains a single Higgs state that 
couples to the other fundamental particles with an intensity proportional 
to their masses \cite{revhixsm}. 
In the case of fermions, the Standard Model
(SM) Higgs couplings are diagonal in flavour 
space, due to the fact that the Higgs couplings and the fermion mass 
matrices are both diagonalised by the same bi-unitary rotations. 
However, this picture ceases to remain valid in many 
extensions of the SM. For instance, in the general 
Two Higgs Doublet Model of Type III (THDM-III), 
where both Higgs doublets couple to both types of up- and
down-type fermions, there appear non-diagonal Higgs couplings, 
which lead to interesting Lepton Flavour Violating/Flavour Changing 
Neutral Current (LFV/FCNC) Higgs phenomenology
\cite{ourTHDM3}. In turn, even though the Minimal Supersymmetric Standard 
Model (MSSM) is a Type II Two-Higgs Doublet Model (THDM-II)
at tree level -- with additional mass and coupling relations enforced
by Supersymmetry (SUSY) --
this structure is not protected by any symmetry, so that
loop effects can effectively render it a THDM-III. 
In addition, the detection of neutrino oscillations
\cite{atmospheric,solar} seem to suggest that there is a large mixing 
between the second and third families in the lepton sector, which could
also appear in new scenarios that are contained in some extensions of 
the SM \cite{DiazCruz:2002er}. In particular, within 
SUSY models, the pattern of LFV effects at the Planck or Grand Unification
Theory (GUT) scales 
could  be reflected in the structure of the soft SUSY-breaking terms, 
i.e., in the slepton mass matrices, which in turn can communicate these 
to the Higgs sector through radiative effects \cite{Babu:1999hn}. 

Detectable effects of LFV Higgs couplings could show up in the decay 
$\tau \to 3\mu$, which is a particularly sensitive
probe at large $\tan\beta$ \cite{Babu:2002et}
(the ratio of the two Higgs vacuum expectation values in the MSSM), 
with a Branching Ratio (BR) scaling as $\tan^6\beta$, a phenomenon which
may render this mode detectable at the LHC.
The relevance of the LFV Higgs decay $\phi\to\tau\mu$\footnote{Hereafter, 
unless otherwise specified, the label $\phi$ will refer to a generic Higgs
state.} 
for Higgs phenomenology at the LHC  was discussed in Ref.
\cite{DiazCruz:1999xe}\footnote{However, a calculation of the actual LFV Higgs
decay rates was presented first in \cite{lfvhpilaftsis}.},  
within the context of several extensions of the SM. 
In particular, it was 
shown that a large BR$(\phi\to \tau\mu)$  (of order 0.001--0.01) 
could easily be achieved in the THDM-III. Moreover. it was shown there
that large LFV Higgs couplings were not in conflict with any low
energy constraints,
such as LFV decays of $\tau$'s. Calculations of the SM Higgs 
BR$(\phi \to \tau\mu)$ showed it to be be very
suppressed ($< 10^{-15}$) whilst in the (constrained) MSSM
 the corresponding rates could be enhanced for some of the Higgs states.

In this paper we are interested in discussing further aspects of 
LFV phenomenology entering the Higgs sector of the MSSM, by
investigating the possibility of a new detection mode for heavy 
$H^0/A^0$ Higgs bosons ($M_{H^0/A^0}\gsim 2M_{W^\pm}$ GeV) at the LHC.
In particular, we shall demonstrate that the decays of the heavy neutral 
Higgs bosons of the MSSM to $\tau \mu$ pairs are sizable and represent 
a very sensitive probe of LFV physics. We calculate the rates for the
BR$(H^0/A^0 \to \tau\mu)$,
involving the heavy neutral MSSM CP-even Higgs state and the CP-odd one,
  and find that they can be as large as $10^{-4}$,
for high values of $\tan\beta$, while those for the BR$(h^0 \to \tau\mu)$,
involving the light neutral MSSM CP-even Higgs state, remain rather small in 
comparison. Furthermore, it should be recalled that
MSSM Higgs production at hadron colliders 
is enhanced for large $\tan\beta$, both via gluon fusion
and in association with $b\bar{b}$ pairs. Moreover,
in this  $\tan\beta$ regime, there  appears a
degeneracy for the masses of the $H^0$ and $A^0$ states,
which essentially doubles the event rate of the overall Higgs
signal. All such effects enable one then to reach detectable levels at the LHC
for $H^0/A^0\to\tau\mu$ signals.
Finally, by extending previous studies on the signal-to-background
separation at the LHC, we show that detection of $H^0/A^0$ LFV Higgs decays
into tau-muon pairs could be achieved for Higgs masses as high as 
600 GeV. 
In short, these LFV decay modes can provide a new detection channel for the 
heavy Higgs bosons of the MSSM, which would in turn give not only important 
evidence for SUSY but also,
along with the modes $B^0\to\mu\mu$, $\tau \to 3\mu$,
$\tau\to\mu\gamma$ and $\mu\to e\gamma$, 
probe the form of the neutrino Yukawa mass matrix.  

The plan of the paper is as follows. 
The underlying aspects of the model and the calculation are laid 
out in Sect.~\ref{sect:LFV}. The LFV Higgs signals are characterised 
in Sect.~\ref{sect:LFVHsignals}. The numerical calculation of the cross 
sections and decays for the relevant Higgs modes
  is pursued in Sect.~\ref{sect:StoB}, 
 including the determination of signal-to-background event rates  
and the proof of detectability of LFV Higgs signals at the LHC. 
Summary and conclusions 
are found in Sect.~\ref{sect:summa}.

\section{Slepton Mixing and LFV in the Higgs Sector}
\label{sect:LFV}

One of the most attractive explanations for the observed neutrino 
masses ~\cite{atmospheric,solar} is the ``see-saw'' mechanism~\cite{seesaw}, 
which includes Dirac masses ($m_D$) as well as Majorana masses ($M_R$). 
Atmospheric neutrino data favours a $\nu_\tau$ mass of about
$0.04\ev$~\cite{nufits}. Thus, for Dirac neutrino masses of the order
of the corresponding up-quark masses, i.e. $(m_D)_{\nu_\tau} \simeq$ $100-200$ 
GeV,
as predicted in a GUT such as SO(10), 
one finds that the  right-handed Majorana mass, $M_R$, needs to
be of order $10^{14}\gev$.  Majorana neutrino masses imply 
LFV within the Minimal Supersymmetric see-saw Model,  
which is defined as the Minimal Supersymmetric Standard Model
augmented by three heavy right-handed neutrinos, $\nu_R$\footnote{Henceforth, 
we will use the notation $\nu$MSSM to 
indicate such an extension of the MSSM.},
LFV interactions can be communicated directly from $\nu_R$'s to the 
sleptons and from these to the charged leptons and Higgs bosons.  The initial 
communication takes place through renormalisation group flow of the slepton 
mass matrices at energies between $\mpl$ and $M_R$.  The presence of $\nu_R$ 
states at scales above $M_R$ leaves an imprint on the mass matrices of the 
sleptons, which propagates down to the Electro-Weak (EW) scale. This effect 
has been used to predict large BRs for $\tau\to\mu\gamma$ 
and $\mu\to e\gamma$ within the MSSM~\cite{gammaold,moroi,gamma}.

To derive the effective Lagrangian for the LFV lepton-Higgs 
interactions, we begin with the Yukawa Lagrangian:
\beq
-{\cal L}=\bar l_R Y_l L_L H_d + \bar\nu_R Y_\nu L_L H_u
+{\textstyle\frac12} \nu_R^\top M_R\, \nu_R,
\eeq
where $l_R$, $L_L$ and $\nu_R$ represent the right-handed charged leptons, 
left-handed lepton doublets and right-handed neutrinos, respectively, while
$H_u,H_d$ denote the Higgs doublets of the MSSM. 
$Y_l$, $Y_\nu$ and $M_R$ are $3\times 3$ matrices in flavour space. 
We shall work in a basis in which both $Y_l$ and $M_R$ have
been diagonalised, but where $Y_\nu$ remains an arbitrary complex matrix.
Lepton number is violated in this Lagrangian due
to the presence of the $\nu_R$ Majorana mass term.

Furthermore, the $6\times6$ slepton mass matrix is written in terms of 
$3\times3$ blocks, as follows:

\beq
M^2_{\Slepton} = \left[
         \begin{array}{ll}
          M_{LL}^2         &  M_{LR}^2\\[1.5mm]
          M_{LR}^{2\,\dag}   &  M_{RR}^2
         \end{array}
         \right],
\eeq
where
\bea
M_{LL}^2 & = & M_{\Slepton}^2+M_l^2+  M_{Z^0}^2 \cos2\beta
\,(T_{3\Slepton} -Q_{\Slepton} \sin^2 \theta_W) \\
M_{RR}^2 & = & M_{\Slepton}^2+M_l^2+   M_{Z^0}^2 \cos2\beta
Q_{\Slepton}\sin^2\theta_W \\
M_{LR}^2 & = & A_l v\,\cos \beta/\sqrt{2}-M_l\,\mu\,\tan\beta
\eea
Here $v=246$ GeV, $\theta_W$ is the weak/Weinberg angle while $\mu$ and $A_l$ 
denote the higgsino mass parameter and trilinear slepton couplings,
respectively.
$M_{W^\pm,Z^0}$ are the masses of the $W^\pm,Z^0$ gauge bosons
and $M_l$ the lepton mass matrix. We will work in 
a basis where $M_l$ is diagonal.

When the SUSY-breaking slepton mass matrix  $(M^2_\Slepton)_{ij}$ evolves from 
the scale $M$ at which flavour-blind SUSY-breaking is communicated 
to the visible sector, down to the slepton mass scale
$M_{\Slepton}$ (assuming $M>M_R$), one obtains 
a flavour-mixing piece that corrects the slepton soft mass terms, i.e.,
$M^2_{\Slepton} \to M^2_{\Slepton} + \Delta M^2_{\Slepton}$,
with the latter given by:
\beq
\left(\Delta M^2_{\Slepton} \right)_{ij} \simeq
-\frac{\log(M/M_R)}{16\pi^2}\left(6m_0^2(Y_\nu^\dagger Y_\nu)_{ij}
+2\left(A_\nu^\dagger A_\nu\right)_{ij}\right),
\eeq
where $m_0$ is a common scalar mass evaluated at the scale $Q=M$ and $i\neq j$. 
If one assumes that the $A$-terms are proportional to Yukawa matrices, 
then:
\beq
\left(\Delta M^2_\Slepton\right)_{ij} \simeq \xi \left(Y_\nu^\dagger
Y_\nu\right)_{ij},
\eeq
where
\beq
\xi =
-\frac{\log(M/M_R)}{16\pi^2}(6+2a^2)m_0^2
\label{log}
\eeq
and $a$ is ${\cal O}(1)$.
In the simplest SUSY-breaking scenarios, gravity plays the
role of messenger and $M=\mpl$.
Global fits to neutrino data favour large mixing between
$\nu_\mu$ and $\nu_\tau$ and also between $\nu_e$ and
$\nu_\mu$~\cite{nufits}. Thus, we shall consider here a form for $m_\nu$ 
with ${\cal O}(1)$ entries in the 23 and 32 elements ~\cite{altarelli}.
If we further assume that $(M_R)_{ij}$ is an identity matrix, then
$(Y_\nu^\dagger Y_\nu)_{23}$ will also be of ${\cal O}(1)$.

This source of LFV interactions can be transmitted to the Higgs sector 
as well, because radiative effects could induce flavour mixing in the Higgs 
couplings.  These corrections allow the neutral Higgs bosons to
mediate  FCNCs, in particular 
$B^0\to\mu\mu$ \cite{Babu:1999hn} can reach BRs at large $\tan\beta$ 
that can be probed by Run~II of the Tevatron~\cite{bmuanalysis}. 
 For the leptonic sector, the Feynman graphs that induce
such corrections involve loops of sleptons and charginos/neutralinos, 
which are transmitted further to induce LFV Higgs couplings.
To derive the SUSY-induced THDM-III, one can write an effective 
Lagrangian for the couplings of the charged leptons to the neutral 
Higgs fields, namely:

\beq
-{\cal L}=\bar l_R Y_l E_{L} H_d^0 +\bar l_R Y_l \left(\eps_1{\bf 1}
+\eps_2 Y_\nu^\dagger Y_\nu\right) E_{L} H_u^{0*} + h.c.
\label{FCLag}
\eeq
where $\epsilon_{1,2}$ include the slepton radiative effects.
LFV couplings results from our inability to simultaneously diagonalise 
the term $Y_l$ and the non-holomorphic loop corrections, 
$\eps_2 Y_l Y_\nu^\dagger Y_\nu$. The contributions from higgsinos and 
gauginos, which are approximated as mass eigenstates, can be written as 
follows, 
\beq
\eps_{2i}\simeq \frac{\alpha_i}{8\pi}\xi\mu M_i f_i
\left(\mu^2, m^2_{\slepton_{La}}, m^2_{\slepton_{Lb}},M_i^2\right),
\label{eq:2b}
\eeq
in which $\slepton_{a}=\smuon, \selectron, \sneutrino_l$, and
$\slepton_{b}=\stau, \sneutrino_{\tau}$. $M_i=M_{1,2}$ are the U(1) 
and SU(2) gaugino masses, while $\xi$ is defined in eq.~(\ref{log}).
For our purposes, the function $f_i$ can be evaluated in the limit
$a=b=c=d$, for which $f_i(a,a,a,a)=1/(6a^2)$.

Since the charged lepton masses cannot be diagonalised in the same 
basis as the Higgs couplings, this will allow neutral Higgs bosons 
to mediate LFV processes, with rates proportional to $\eps_2^2$. 
The term proportional to $\eps_1$ will generate a mass shift for the 
charged leptons that will appear as a second-order effect \cite{mysuperadfm}.

Then, the LFV interactions relevant for the Higgs sector phenomenology
can be written in terms of Higgs mass eigenstates as follows 
($\phi^0_k=h^0,H^0,A^0$):
\beq
-{\cal L}_{\phi_k l_il_j} = [\frac{g\, m_\tau \eta_\phi}{2M_{W^\pm} \cos\beta}] 
\left( \lambda^{\phi_k}_{ij}  \bar\l_{Ri}\,\l_{Lj} \phi_k^0 + h.c.\right),
\label{finalL}
\eeq
where 
\beq
\lambda^{\phi_k}_{ij}\simeq 
\frac{- \eps_2 \tan\beta \rho_\phi \left(Y_\nu^\dagger Y_\nu\right)_{ij} }
     {\sin\beta \eta_\phi}
\eeq
and $(\eta_h, \eta_H, \eta_A)=(\sin\alpha,-\cos\alpha,\sin\beta)$,
$(\rho_h, \rho_H, \rho_A)= (\cos(\beta-\alpha),-\sin(\beta-\alpha),-i)$,
with $\alpha$ denoting the Higgs mixing angle.
Constraints on the LFV $(\bar\tau_R\, \mu_L)$--Higgs interaction 
can be obtained from the LFV $\tau$-decays (e.g., $\tau\to3\mu$), which
can be generated via exchange of $h^0$, $H^0$ and $A^0$. For instance, for
the case in which 
$\mu= M_1= M_2= m_{\slepton}=
m_{\sneutrino}$, $M_R=10^{14}\gev$ and $(Y_\nu^\dagger Y_\nu)_{32}=1$,
Ref.~\cite{Babu:2002et} finds that $\eps_2 \simeq 4\times 10^{-4}$, which is  
stable with respect to changes in the SUSY spectrum. Then
$\mbox{BR}(\tau\to3\mu) \simeq (1\times 10^{-7})\times(\tan\beta/60)^6
\times(100\gev/M_{A^0})^4$, which puts the $\tau\to3\mu$ mode into 
a regime that is experimentally accessible at current B-factories. 
At the LHC and SuperKEKB, limits in the region of $10^{-9}$ should be 
achievable~\cite{exp}, allowing an even deeper probe into the model
parameter space. On the other hand, Tevatron has already constrained 
the large $\tan\beta$ domain, from the search for the decay 
$B_s\to \mu\mu$ \cite{TevatronBsmumu}, that seems to exclude the 
value $\tan\beta=60$, which is preferred by the requirement of
Yukawa unification. However, it is possible to evade such constraints,
for instance in SUSY breaking scenarios where slepton and squark
masses do not have any strong correlation. In any case, given that
such limits depend on multiple MSSM parameters, which makes it difficult 
to draw a general conclusion, it is certainly preferable to test the
resulting LFV Higgs couplings directly at the LHC, as it is discussed
in the next section.

\section{The LFV Higgs Decays  $H^0/A^0 \to\tau\mu$} 
\label{sect:LFVHsignals}
\begin{figure}[!t]
\begin{center}
\vspace{-7.cm}
{\epsfig{file=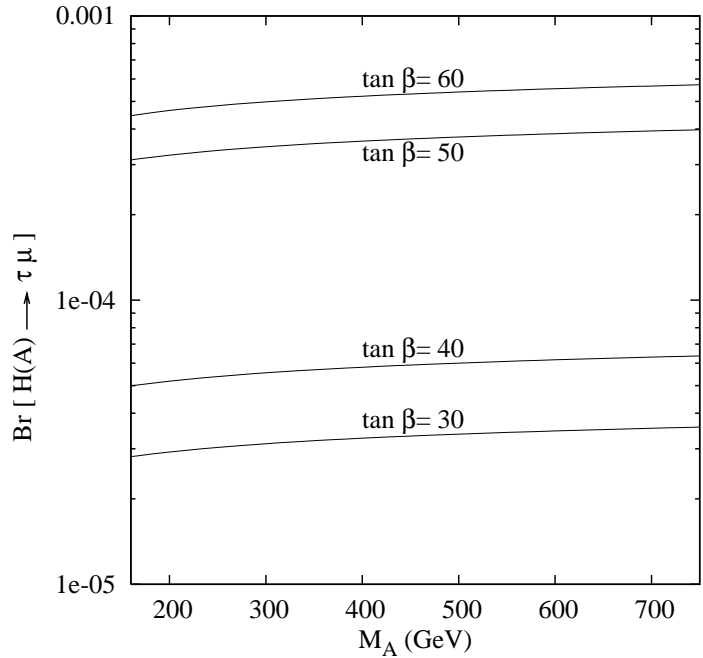, width=9.9cm, angle=0}}
\end{center}
\vspace*{-0.75cm}
\caption{\small The variation of the LFV BR of heavy Higgs 
bosons $(H^0, A^0)$ as a function of $M_{A^0} ( M_{H^0} = M_{A^0})$ for 
for $\tan\beta = 30, 40$ (with set A) and 50,60 (with set B).}
\label{fig:BRs}
\end{figure}

 LFV Higgs decays have been evaluated within the general MSSM with a 
particular ansatz for the trilinear $A$ terms in  \cite{DiazCruz:2002er}, 
where was found that a BR$(h^0 \to \tau\mu) \simeq10^{-4}-10^{-7}$ could 
well be achieved.  
Subsequently, Refs. \cite{Brignole:2003iv,Brignole:2004ah}
and \cite{Arganda:2004bz} presented a more detailed calculation of the  
BR$(h^0 \to \tau\mu)$, within the MSSM, which was essentially in agreement 
with the previous result. In fact,  Ref. \cite{Arganda:2004bz} also reported  
a complete one-loop calculation of the LFV Higgs decay within the SM 
extension with massive neutrinos, using a realistic pattern of neutrino 
masses and mixings, which resulted in  very suppressed LFV Higgs decays 
(with BR of order $10^{-30}$ or less).  Afterwards, \cite{Han:2000jz}
presented a detailed study of the prospects to detect LFV Higgs
decays at the Tevatron and LHC, concluding that it is
certainly possible to detect such decays in the THDM-III
within the Higgs mass range 114--160 GeV, approximately.
Later on, Ref. \cite{Assamagan:2002kf} presented a more realistic
study of the signal and backgrounds, essentially reaching the same
conclusions.  Subsequently, it was also studied in detail the mass-matrix
ansatz used in the THDM-III, by Ref.~\cite{DiazCruz:2004pj},
while the evaluation of the corresponding LFV Higgs decays 
was presented in \cite{GomezBock:2005hc}.
Mixing of the SM fermions with other exotic fermions
was also shown to be a possible source of LFV in
the Higgs sector \cite{Cotti:2002zq},
resulting in BRs of the order 0.01-0.001 again,
which could clearly be detectable too.
Bounds on  LFV Higgs decays at the Tevatron were reported
by the CDF collaboration in \cite{Acosta:2003jd}.

In this paper we shall concentrate on the  LFV heavy Higgs decays to $\tau \mu$, 
which has a very small BR within the context of the SM
 with light neutrinos, so that this channel is potentially an excellent 
window for probing new physics. Although the previously mentioned works
have studied LFV Higgs decays within the $\nu$MSSM, the specific evaluation
of the BRs of the LFV decays for $H^0/A^0$ in the heavy mass range has not
been studied. Hence a discussion of the corresponding detectability at 
the LHC has not been presented so far either. To remedy this is our aim in 
the present paper.

In order to derive the formulae for the LFV Higgs decay widths, we
notice that the quantity inside the square brackets in eq. (11)
corresponds to the Higgs-lepton coupling $\phi_k \tau \bar{\tau}$, 
which will be denoted by $g_{\phi \tau \bar{\tau}}$. 
Thus, we can write the LFV Higgs coupling $\phi_k \tau \mu$ as
\beq
g_{\phi \tau \mu}= g_{\phi \tau \bar{\tau}} \lambda^\phi_{ij}.
\eeq
The decay width for the generic process $\phi_k \to \tau \mu$ 
(in which we add both final states $\tau^+ \mu^-$ and $\tau^- \mu^+$)
can then be written in terms of the Higgs decay width 
$\Gamma (\phi \to \tau \tau)$, as follows:
\beq\label{eq:width}
\Gamma (\phi\to \tau \mu) \, =\,2 |\lambda^\phi_{\tau\mu}|^2
                                   \Gamma (\phi \to \tau \tau),
\eeq
so that the LFV Higgs BR can in turn be approximated by
BR$(\phi \to \tau \mu)= 2 |\lambda^\phi_{\tau\mu}|^2 
{\rm{BR}}(\phi \to \tau \tau)$.

We are interested in studying the large $\tan\beta$ domain
 (i.e., $\beta \to \pi/2$), where the LFV Higgs couplings are enhanced.
It is also simpler to work in such so-called decoupling regime of the
 MSSM Higgs sector,
which in fact is quite general since it is reached even for moderate 
values of $M_{A^0}$ ($\simeq 200$ GeV). In this case we have that 
$\lambda^\phi_{ \tau\mu} \to 0, \epsilon_2 \tan\beta, \epsilon_2 \tan\beta$
for $\phi_k=h^0,H^0,A^0$, respectively.
Therefore,  the LFV decays of the light Higgs boson
($h^0$) are suppressed for most regions of parameter space.
Conversely, for the above mentioned choices of SUSY parameters yielding
$\epsilon_2=4\times 10^{-4}$ (which we call set A)  and with 
$M_{H^0,A^0}\approx 160$ GeV, one obtains
BR$(H^0/A^0 \to \tau \tau) \simeq  0.12$,  
which in turn gives
BR$(H^0/A^0 \to \tau \mu) \simeq  2.9 \times 10^{-5}$ 
for $\tan\beta=30$.
For another set of parameters with a large $\mu$ limit, i.e.
$\mu >> M_{1,2}$ (which we call set B),  one gets  
$\epsilon_2 \simeq 8 \times 10^{-4}$,
which will produce a larger BR for $H^0/A^0 \to \tau \mu$. 
More in general, we  calculate the LFV Higgs decay rates in the channels
$H^0/A^0\to\tau\mu$ -- by appropriately 
modifying the {\tt HDECAY} program \cite{Djouadi:1997yw}
 and using the formula in 
eq.~(\ref{eq:width}) -- as a function of the Higgs masses and
corresponding LFV Higgs couplings. 


It is appropriate to mention at this point that reference \cite{Assamagan:2002kf}
did include a discussion of the LFV $H^0/A^0$ decays. However, the
authors concentrated
on the mass range below $2M_{W^\pm}\approx160$ GeV. They hint in fact that above this
mass range the modes $H^0/A^0\to \tau\mu$ will be suppressed because the 
channels
$W^+W^-$ and $Z^0Z^0$ would be open and dominate the total decay width. However, this 
is not true. The reason is twofold. Firstly, the $A^0$ -- being a CP-odd state -- does not couple to vector 
boson pairs. Secondly, although the $H^0$ state does couple to $W^+W^-$ and $Z^0Z^0$ pairs, in the
heavy Higgs mass limit such coupling is considerably suppressed. Our calculation
takes correct care of these aspects.

\section{Signal-to-Background Analysis}
\label{sect:StoB}
\begin{figure}[!t]
\begin{center}
\vspace{-7.cm}
{\epsfig{file=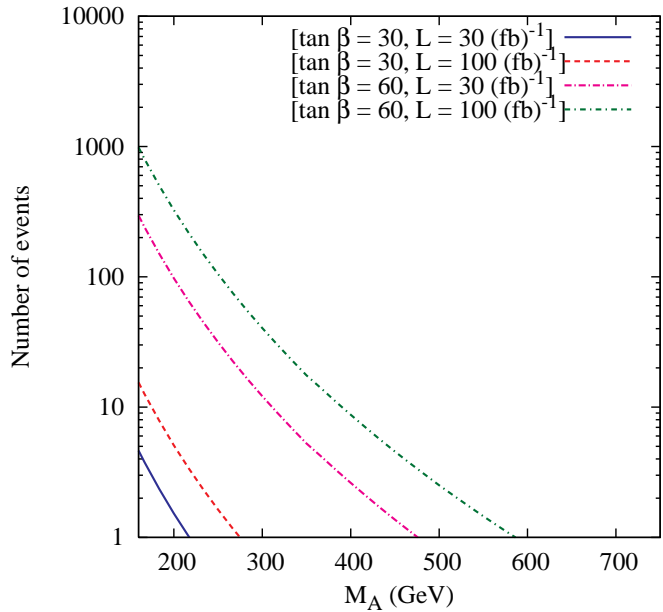, width=9.5cm, angle=0}}
\end{center}
\vspace*{-0.75cm}
\caption{\small The dependence  of number of events of the LFV Higgs
signals at the LHC on the degenerate Higgs masses 
$M_{H^0}$ and $M_{A^0}$, including the
sum of production cross sections $\sigma(gg\to H^0) +
\sigma(gg\to A^0)$ 
times the corresponding decay rates
BR$(H^0\to\tau\mu)$ and BR$(A^0\to\tau\mu)$, for two representative
values of $\tan\beta = 30$ (+set A)  and 60 (+set B). 
We are assuming a detection efficiency 
of 3\% and two values of integrated luminosities $30~{\rm fb}^{-1}$
and $100~{\rm fb}^{-1}$.}
\label{fig:events-ggHA}
\end{figure}
\begin{figure}[!t]
\begin{center}
\vspace{-7.cm}
{\epsfig{file=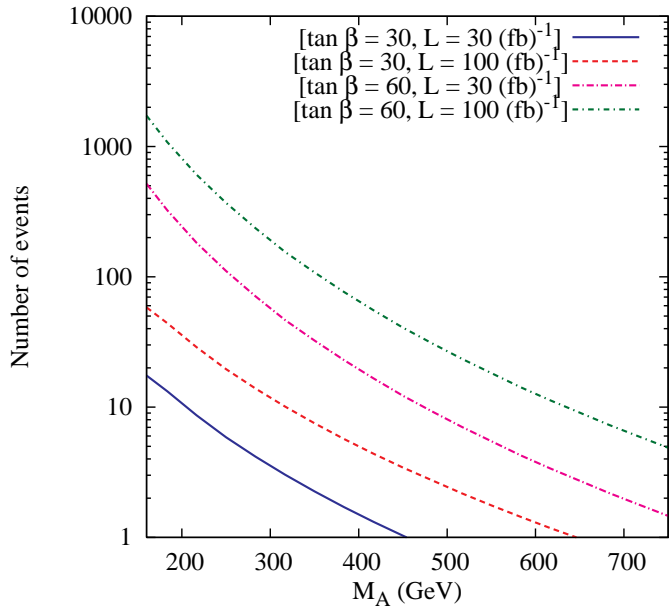, width=9.5cm, angle=0}}
\end{center}
\vspace*{-0.75cm}
\caption{\small The dependence  of number of events of the LFV Higgs
signals at the LHC on the degenerate Higgs masses 
$M_{H^0}$ and $M_{A^0}$, including the
sum of production cross sections 
$\sigma(pp\to b {\bar b}H^0 ) + \sigma(pp\to  b {\bar b}A^0 )$ 
times the corresponding decay rates
BR$(H^0\to\tau\mu)$ and BR$(A^0\to\tau\mu)$. Other parameters are the same 
as in Figure \ref{fig:events-ggHA}.}
\label{fig:events-HAbb}
\end{figure}

Once one folds in the values of LFV Higgs BRs with the 
main heavy neutral Higgs production modes at the LHC, 
it is clear that LFV Higgs decays into $\tau\mu$ pairs may be
detectable at the CERN hadron collider. As a benchmark, 
according to previous studies \cite{Assamagan:2002kf}, with SM-like cross 
sections and $m_\phi \lsim 160$ GeV, one could detect at the LHC
the aforementioned LFV Higgs decays with a BR of order $8 \times 10^{-4}$.
We intend to push forward the region of detectability into higher
mass values
within the MSSM, by 
exploiting the aforementioned $\tan\beta$ enhancement and the fact that
our LFV Higgs BRs become almost constant for heavier Higgs masses. 
Besides, for heavier Higgs masses one
should expect a much larger background reduction, as compared to the
lower Higgs mass case, owning to the much harder energy spectra for 
the emerging $\tau$- and $\mu$-leptons. 

As already mentioned, high values of $\tan\beta$ are also associated
with a large $b$-quark Yukawa coupling, which in turn can produce
an enhancement of the Higgs production cross sections at hadron
colliders via both gluon fusion and associated production with
$b$-quark pairs: i.e., $gg \to H^0/A^0$ (via triangle loops at lowest order) 
and $gg,q\bar q\to b\bar bH^0/A^0$ (at tree-level), respectively.
(We calculate these two production processes here by using  the {\tt HIGLU}
and {\tt HQQ} programs in default configurations~\cite{Spira:1995mt}.)
Figure~\ref{fig:BRs} shows the relevant LFV Higgs BRs as
a function of $M_{A^0}$. The LFV BR is basically the same for both $H^0$ and
$A^0$ and its variation with $M_{A^0}$ is very mild. 
In Figures ~\ref{fig:events-ggHA} and \ref{fig:events-HAbb} 
we present the expected LFV Higgs event rates as a function of degenerate 
Higgs mass $M_{A^0} $ for two values of $\tan\beta = 30 $ and 60 and 
two values of LHC luminosities,
 $L = 30~{\rm fb}^{-1}$ and $100~{\rm fb}^{-1}$,
respectively. It is clear from these two plots that LFV Higgs rates
con be substantial even at large Higgs masses.

In Ref.~\cite{Han:2000jz} it was proposed a series of cuts to reconstruct
the hadronic and electronic $\tau$  decays from $\phi\to \tau\mu$ and separate the
signal from the background, which is dominated by $\tau$-pair production
via Drell-Yan modes (i.e., $q\bar q\to \gamma^*,Z^{(*)}\to \tau^+\tau^-$)
and $q\bar q,gg\to W^+W^-\to \tau^+\nu_\tau\tau^-\bar\nu_\tau$.
In fact, it should be
recalled that the decay product distributions of $\tau$-leptons
generated in the decay of Higgs bosons are notably different
from those emerging in gauge boson decays, because of the
different spin of the primary objects. 
A more realistic search strategy for the LHC based on the cuts
of  Ref.~\cite{Han:2000jz} was
presented in Ref.~\cite{Assamagan:2002kf}, where one can find
detection efficiencies in the Higgs mass range 120--160 GeV. 
The typical figure goes from
about 2\% for $M_{h^0} =120$ GeV up to about  3\% for $M_{h^0}=160$ GeV, where
it starts stabilising.
Although one expects that this detection efficiency will increase for
heavier Higgs masses, in order to use a conservative estimate, we shall use 
the 3\% figure throughout in our estimates in the remainder of the paper.


\begin{figure}[!t]
\begin{center}
\vspace{-5.cm}
{\epsfig{file=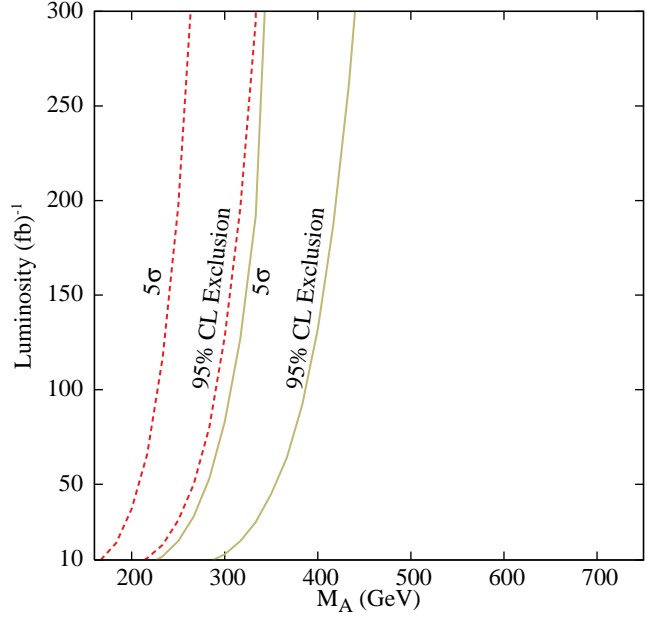, width=9.cm, angle=0}}
\end{center}
\vspace*{-0.75cm}
\caption{\small The 95\% CL exclusion and $5\sigma$ discovery reaches
for the LFV Higgs
signals at the LHC as a function of the degenerate Higgs masses 
$M_{H^0}$ and $M_{A^0}$
and  the integrated luminosity, including
the sum of production cross sections $\sigma(gg\to H^0)$ and 
$\sigma(gg\to A^0)$ plus $\sigma(pp\to b {\bar b}H^0 )$
 and $\sigma(pp\to b {\bar b}A^0 )$
times the corresponding decay rates
BR$(H^0\to\tau\mu)$ and BR$(A^0\to\tau\mu)$, for 
$\tan\beta =$ 30 (+set A) (dashed lines) 
and 40 (+set A) (solid lines).}
\label{fig:reach30-40}
\end{figure}

\begin{figure}[!t]
\begin{center}
\vspace{-5.cm}
{\epsfig{file=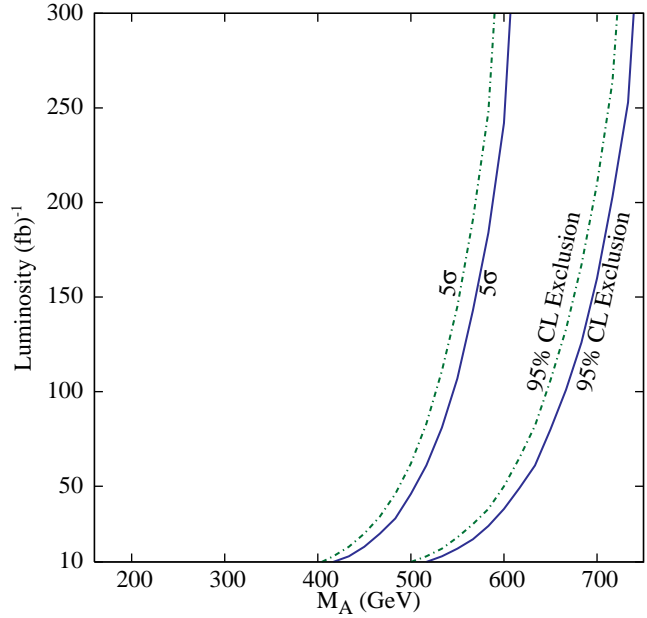, width=9.cm, angle=0}}
\end{center}
\vspace*{-0.75cm}
\caption{\small Same as Figure \ref{fig:reach30-40}, but for 
$\tan\beta =$ 50 (+set B) (dotted-dashed lines) and 60 (+set B) (solid lines).}
\label{fig:reach50-60}
\end{figure}

Adopting the background rates estimated in Ref.~\cite{Assamagan:2002kf}
for Higgs masses up to 200 GeV and trivially extrapolating them
to heavier Higgses, we are then in a position to compare the yield of our 
signals (see Figures~\ref{fig:events-ggHA}--\ref{fig:events-HAbb})
with that of the total background, thereby estimating both a 95\% Confidence
Level (CL) exclusion limit and a $5\sigma$ discovery reach. 
The scope of the LHC in both respects, as a function of the Higgs 
masses and machine luminosity, is then well described by
Figures~\ref{fig:reach30-40} and \ref{fig:reach50-60}. To display our results,
we have choosen the combinations $\tan\beta=30$ and $40$ 
with SUSY parameters of set A, and  $\tan\beta=50$ and $60$ 
with SUSY parameters of set B.
(Other combinations should lay within these results.)
Clearly, with an integrated luminosity of 300 fb$^{-1}$,
for $\tan\beta=30$, one can detect
a signal for Higgs masses up to about 260 GeV, which is already 
significantly above the $2M_{W^\pm}$ mark  of Ref.~\cite{Assamagan:2002kf}, 
while, for  $\tan\beta=60$,  Higgs masses up to even 600 GeV can be probed. 
However, since for $\tan\beta=60$ one is dangerously close to the  bounds 
from $B\to \mu \mu$, the reader may well refer instead
to those for $\tan\beta=40$ (with SUSY parameters of set A) and
$\tan\beta=50$ (with SUSY parameters of set B). 
In these cases too we find that it will be possible to extract
a LFV Higgs signal at the LHC for heavy Higgs masses, up to about
300(500) GeV for  $\tan\beta=40$(50) with 300 fb$^{-1}$ of luminosity.

\section{Summary and Conclusions}
\label{sect:summa}

We have demonstrated that LFV effects in the slepton sector
of the $\nu$MSSM can generate LFV couplings 
between leptons and neutral Higgs bosons
leading to large BRs for LFV Higgs decays into lepton pairs.
In particular, we have calculated the BRs of the processes
$H^0/A^0 \to \tau\mu$ and found that they can be as large as 
$3 \times 10^{-4}$, 
while the BR$(h^0 \to \tau\mu)$ is only about $\simeq 10^{-8}$. Furthermore, 
these rates occur for large values of $M_{H^0/A^0}$ and
$\tan\beta$, a configuration also responsible for 
a strong degeneracy between the masses (and couplings) of the $H^0$ and 
$A^0$ states, producing an overall Higgs event rate which is double
the one of either Higgs state alone. These LFV Higgs modes can be extracted at
the LHC for Higgs masses slightly beyond 600 GeV, provided
$\tan\beta=60$. For smaller values of this parameter though the LHC
scope greatly diminishes, reducing to just above 260 GeV in Higgs mass for $\tan\beta=30$. 
These values  can only be reached at 300 fb$^{-1}$ of luminosity. However, even
with a modest 10 fb$^{-1}$, one could probe Higgs masses up to 400(415) GeV, provided
$\tan\beta=50$(60).
Besides, in view of the assumptions made on detection efficiencies for the
signal, we believe these conclusions to be rather conservative.
Altogether, these novel channels complement the modes
$B^0\to\mu\mu$, $\tau \to 3\mu$, $\tau\to\mu\gamma$ and $\mu\to e\gamma$
in order to provide evidence for SUSY and key insights 
into the form of the neutrino Yukawa mass matrix. More detailed experimental
simulations are now awaited.
\vskip0.25cm\noindent
{\sl Acknowledgments}~We would like to thank the CERN Theory Unit for their
hospitality when this work was started and K. Babu and R. Martinez
for discussions. This research was supported in part by CONACYT-SNI 
(M\'exico) and STFC (UK). The work of DKG was partially supported by the
Department of Science \& Technology (DST), Govt. of India, project No. 
SR/S2/HEP-12/2006. DKG also thanks the hospitality provided by the 
Regional Centre for Accelerator Based Particle Physics, HRI, Allahabad, 
India, where part of this work was done.

\end{document}